\RequirePackage{snapshot}
\pdfoutput=1
\documentclass[a4paper,11pt]{article}

\usepackage{url}

\usepackage{fullpage}

\usepackage{amsmath,amsthm,amssymb}
\usepackage{xcolor}
\usepackage{graphicx}
\usepackage{bbold}
\usepackage{textcomp} 
\usepackage{siunitx}

\usepackage{listings}
\usepackage{courier}

\usepackage[position=top]{subfig}
\usepackage{authblk}
\renewcommand\Affilfont{\itshape\small}

\title{A full-fledged micromagnetic code in less than 70 lines of NumPy}

\author[1]{Claas Abert\thanks{claas.abert@tuwien.ac.at}}
\author[1]{Florian Bruckner}
\author[2]{Christoph Vogler}
\author[1]{Roman Windl}
\author[1]{Raphael Thanhoffer}
\author[1]{Dieter Suess}
\affil[1]{Christian Doppler Laboratory of Advanced Magnetic Sensing and Materials, Institute of Solid State Physics, Vienna University of Technology, Austria}
\affil[2]{Institute of Solid State Physics, Vienna University of Technology, Austria}

\begin{document}
\lstset{
language=Python,
frame=single,
basicstyle=\footnotesize\ttfamily
}

\maketitle

\begin{abstract}
  We present a complete micromagnetic finite-difference code in less than 70 lines of Python.
  The code makes largely use of the NumPy library and computes the exchange field by finite differences and the demagnetization field with a fast convolution algorithm.
  Since the magnetization in finite-difference micromagnetics is represented by a multi-dimensional array and the NumPy library features a rich interface for this data structure, the presented code is a good starting point for the development of novel algorithms.

\end{abstract}
  
\newpage
\section{Introduction}
Micromagnetic simulations have become an integral tool for the investigation of ferromagnetic nano structures.
A lot has been published on algorithms and programming paradigms used for the numerical solution of the micromagnetic equations and a variety of open and closed source micromagnetic codes is available.
These codes can be roughly divided into those acting on regular cuboid grids \cite{donahue1999oommf,abert2012fast,vansteenkiste2014design} and those acting on irregular grids \cite{abert2013magnum,fischbacher2007systematic,scholz2003scalable,suess2002time,kakay2010speedup,chang2011fastmag}.
Irregular-grid codes usually employ finite-elements or fast-multipole methods for spatial discretization \cite{schrefl2007numerical}.
A popular class of regular grid methods applies the finite-difference method for the computation of the exchange field and a fast convolution for the computation of the demagnetization field \cite{miltat2007numerical}.
In this work we present a complete micromagnetic code of the latter kind that is written in only 70 lines of Python and makes largely use of the NumPy library \cite{van2011numpy}.

The presented code is not able to compete with mature finite-difference codes in terms of performance and flexibility.
However, it delivers all essential building blocks of a micromagnetic code and is therfore perfectly suited for prototyping new micromagnetic algorithms.
In particular, the NumPy library is a good choice for the presented code since the magnetization in finite-difference micromagnetics is represented by an $n$-dimensional array and the NumPy library has a very powerful interface for $n$-dimensional arrays that supports a large variety of operations.

The paper is structured as follows.
Section~\ref{sec:model} gives a brief overview over the micromagnetic model.
In Sec.~\ref{sec:demag}-\ref{sec:llg} the implementation of the integral micromagnetic subproblems is described.
The presented code is validated by numerical experiments in Sec.~\ref{sec:experiments}.

\section{Micromagnetic Model}\label{sec:model}
The central equation of dynamic micromagnetics is the Landau-Lifshitz-Gilbert equation that describes the motion of a continuous magnetization configuration $\boldsymbol{m}$ in an effective field $\boldsymbol{H}_\text{eff}$
\begin{equation}
  \frac{\partial \boldsymbol{m}}{\partial t} =
  - \frac{\gamma}{1 + \alpha^2} \boldsymbol{m} \times \boldsymbol{H}_\text{eff}
  - \frac{\alpha \gamma}{1 + \alpha^2} \boldsymbol{m} \times (\boldsymbol{m} \times \boldsymbol{H}_\text{eff})
  \label{eq:llg}
\end{equation}
where $\gamma$ is the reduced gyromagnetic ratio and $0 \leq \alpha \leq 1$ is a dimensionless damping constant.
The effective field $\boldsymbol{H}_\text{eff}$ is given by the negative variational derivative of the free energy
\begin{equation}
  \boldsymbol{H}_\text{eff} = - \frac{1}{\mu_0 M_\text{s}} \frac{\delta U}{\delta \boldsymbol{m}}
  \label{eq:effective_field}
\end{equation}
where $\mu_0$ is the magnetic constant and $M_\text{s}$ is the saturation magnetization.
Contributions to the effective field usually include the demagnetization field, the exchange field, the external Zeeman field and terms describing anisotropic effects.
In this work we focus on the numerical computation of the demagnetization field, see Sec.~\ref{sec:demag}, and the exchange field, see Sec.~\ref{sec:exchange} as well as the integration of the Landau-Lifshitz-Gilbert equation, see Sec.~\ref{sec:llg}.

For the numerical solution of \eqref{eq:llg} and \eqref{eq:effective_field} a regular cuboid grid is used for the spatial discretization.
Every simulation cell is of size $\Delta r_1 \times \Delta r_2 \times \Delta r_3$ and can be addressed by a multiindex $\boldsymbol{i} = (i_1, i_2, i_3)$.
All spatially varying quantities such as the magnetization $\boldsymbol{m}$ are thus represented by an $n$-dimensional array
\begin{equation}
  \boldsymbol{m}(\boldsymbol{r}) \approx
  \boldsymbol{m}_{\boldsymbol{i}}.
\end{equation}
The Python library NumPy provides the class \texttt{ndarray} for this purpose that supports a large number of operations.
Despite Python being a scripting language, all collective operations of \texttt{ndarray} have a good performance due to the native implementaion of the NumPy library.

\section{Demagnetization Field}\label{sec:demag}
The demagnetization field accounts for the dipole-dipole interaction of the elementary magnets.
For a continuous magnetization configuration the demagnetization field is given by
\begin{align}
  \boldsymbol{H}_\text{demag}(\boldsymbol{r}) &= M_\text{s} \int_\Omega \boldsymbol{\tilde{N}}(\boldsymbol{r} - \boldsymbol{r}') \boldsymbol{m}(\boldsymbol{r}') \,\text{d}\boldsymbol{r}' \\
  \boldsymbol{\tilde{N}}(\boldsymbol{r} - \boldsymbol{r}') &= - \frac{1}{4 \pi} \boldsymbol{\nabla} \boldsymbol{\nabla}' \frac{1}{| \boldsymbol{r} - \boldsymbol{r}'|}
\end{align}
where $\boldsymbol{\tilde{N}}$ is called demagnetization tensor.
This expression has the form of a convolution of the magnetization $\boldsymbol{m}$ with the matrix valued kernel $\boldsymbol{\tilde{N}}$.
By choice of a regular grid this convolution structure can also be exploited on the discrete level
\begin{align}
  \boldsymbol{H}_{\boldsymbol{i}} &=
  M_\text{s} \sum_{\boldsymbol{j}} \boldsymbol{\tilde{N}}_{\boldsymbol{i} - \boldsymbol{j}} \boldsymbol{m}_{\boldsymbol{j}}
  \label{eq:demag_convolution_discrete} \\
  \boldsymbol{\tilde{N}}_{\boldsymbol{i} - \boldsymbol{j}} &=
  \frac{1}{\Delta r_1 \Delta r_2 \Delta r_3} \iint_{\Omega_\text{cell}} \boldsymbol{\tilde{N}} \left(
  \sum_k (i_k - j_k) \Delta r_k \boldsymbol{e}_k + \boldsymbol{r} - \boldsymbol{r}'
  \right) \,\text{d}\boldsymbol{r} \,\text{d}\boldsymbol{r}'
  \label{eq:demag_tensor_discrete}
\end{align}
where $\Omega_\text{cell}$ describes a cuboid reference cell and $\boldsymbol{e}_k$ is a unit vector in direction of the $k$th coordinate axis.
Here the magnetization is assumed to be constant within each simulation cell and the field generated by each source cell is averaged over each target cell.
This results in a sixfold integral for the computation of the discrete demagnetization tensor $\boldsymbol{\tilde{N}}_{\boldsymbol{i} - \boldsymbol{j}}$.
An analytical solution of \eqref{eq:demag_tensor_discrete} was derived by Newell et al. in \cite{newell1993generalization}.
The diagonal element $N^{1,1}$ computes as
\begin{multline}
  N_{\boldsymbol{i} - \boldsymbol{j}}^{1,1} =  - \frac{1}{4 \pi \Delta r_1 \Delta r_2 \Delta r_3}
  \sum_{\boldsymbol{k}, \boldsymbol{l} \in \{0,1\}}
  (-1)^{\sum_x k_x + l_x} \\
  f[(i_1 - j_1 + k_1 - l_1) \Delta r_1, (i_2 - j_2 + k_2 - l_2) \Delta r_2, (i_3 - j_3 + k_3 - l_3) \Delta r_3]
  \label{eq:demag_n11}
\end{multline}
where the auxiliary function $f$ is defined by
\begin{align}
  f(r_1, r_2, r_3)
  &= \frac{|r_2|}{2} (r_3^2 - r_1^2) \sinh^{-1}\left( \frac{|r_2|}{\sqrt{r_1^2 + r_3^2}} \right) \nonumber \\
  &+ \frac{|r_3|}{2} (r_2^2 - r_1^2) \sinh^{-1}\left( \frac{|r_3|}{\sqrt{r_1^2 + r_2^2}} \right) \nonumber \\
  &- |r_1 r_2 r_3| \tan^{-1} \left( \frac{|r_2 r_3|}{r_1 \sqrt{r_1^2 + r_2^2 + r_3^2}} \right) \nonumber \\
  &+ \frac{1}{6} (2 r_1^2 - r_2^2 - r_3^2) \sqrt{r_1^2 + r_2^2 + r_3^2}.
\end{align}
The elements $N^{2,2}$ and $N^{3,3}$ are obtained by circular permutation of the coordinates
\begin{align}
  N_{\boldsymbol{i} - \boldsymbol{j}}^{2,2} &= N_{(i_2, i_3, i_1) - (j_2, j_3, j_1)}^{1,1} \\
  N_{\boldsymbol{i} - \boldsymbol{j}}^{3,3} &= N_{(i_3, i_1, i_2) - (j_3, j_1, j_2)}^{1,1}
\end{align}
According to Newell the off-diagonal element $N^{1,2}$ is given by
\begin{multline}
  N_{\boldsymbol{i} - \boldsymbol{j}}^{1,2} = - \frac{1}{4 \pi \Delta r_1 \Delta r_2 \Delta r_3}
  \sum_{\boldsymbol{k}, \boldsymbol{l} \in \{0,1\}}
  (-1)^{\sum_x k_x + l_x} \\
  g[(i_1 - j_2 + k_1 - l_1) \Delta r_1, (i_2 - j_2 + k_2 - l_2) \Delta r_2, (i_3 - j_3 + k_3 - l_3) \Delta r_3]
  \label{eq:demag_n12}
\end{multline}
where the function $g$ is defined by
\begin{align}
  g(r_1, r_2, r_3)
  &= (r_1 r_2 r_3) \sinh^{-1}\left( \frac{r_3}{\sqrt{r_1^2 + r_2^2}} \right) \nonumber \\
  &+ \frac{r_2}{6} (3 r_3^2 - r_2^2) \sinh^{-1}\left( \frac{r_1}{\sqrt{r_2^2 + r_3^2}} \right) \nonumber \\
  &+ \frac{r_1}{6} (3 r_3^2 - r_1^2) \sinh^{-1}\left( \frac{r_2}{\sqrt{r_1^2 + r_3^2}} \right) \nonumber \\
  &- \frac{r_3^3}{6} \tan^{-1} \left( \frac{r_1 r_2}{r_3 \sqrt{r_1^2 + r_2^2 + r_3^2}} \right)
   - \frac{r_3 r_2^2}{2} \tan^{-1} \left( \frac{r_1 r_3}{r_2 \sqrt{r_1^2 + r_2^2 + r_3^2}} \right) \nonumber \\
  &- \frac{r_3 r_1^2}{2} \tan^{-1} \left( \frac{r_2 r_3}{r_1 \sqrt{r_1^2 + r_2^2 + r_3^2}} \right)
   - \frac{r_1 r_2 \sqrt{r_1^2 + r_2^2 + r_3^2}}{3}.
\end{align}
Again other off-diagonal elements are obtained by permutation of coordinates
\begin{align}
  N_{\boldsymbol{i} - \boldsymbol{j}}^{1,3} &= N_{(i_1, i_3, i_2) - (j_1, j_3, j_2)}^{1,2} \\
  N_{\boldsymbol{i} - \boldsymbol{j}}^{2,3} &= N_{(i_2, i_3, i_1) - (j_2, j_3, j_1)}^{1,2}.
\end{align}
The remaining components of the tensor are obtained by exploiting the symmetry of $\boldsymbol{\tilde{N}}$, i.e. $N^{i,j} = N^{j,i}$.

\begin{lstlisting}[caption={Definition of auxiliary functions $f$ and $g$. The very small number \texttt{eps} is added to denominators in order to avoid division-by-zero errors.},label=lst:demag_fg]
eps = 1e-18

def f(p):
  x, y, z = abs(p[0]), abs(p[1]), abs(p[2])
  return + y/2.0*(z**2-x**2)*asinh(y/(sqrt(x**2+z**2)+eps)) \
         + z/2.0*(y**2-x**2)*asinh(z/(sqrt(x**2+y**2)+eps)) \
         - x*y*z*atan(y*z/(x * sqrt(x**2+y**2+z**2)+eps))   \
         + 1.0/6.0*(2*x**2-y**2-z**2)*sqrt(x**2+y**2+z**2)

def g(p):
  x, y, z = p[0], p[1], abs(p[2])
  return + x*y*z*asinh(z/(sqrt(x**2+y**2)+eps))                 \
         + y/6.0*(3.0*z**2-y**2)*asinh(x/(sqrt(y**2+z**2)+eps)) \
         + x/6.0*(3.0*z**2-x**2)*asinh(y/(sqrt(x**2+z**2)+eps)) \
         - z**3/6.0  *atan(x*y/(z*sqrt(x**2+y**2+z**2)+eps))    \
         - z*y**2/2.0*atan(x*z/(y*sqrt(x**2+y**2+z**2)+eps))    \
         - z*x**2/2.0*atan(y*z/(x*sqrt(x**2+y**2+z**2)+eps))    \
         - x*y*sqrt(x**2+y**2+z**2)/3.0
\end{lstlisting}
The calculation of the discrete demagnetization tensor in Python is straightforward.
Listing~\ref{lst:demag_fg} shows the function definitions for the auxiliary functions $f$ and $g$.
Note that fractions occuring in $f$ and $g$ might feature zero denominators.
However, limit considerations show that all fractions tend to zero in this case.
In order to avoid devision-by-zero errors in the implementation, a very small floating point number \texttt{eps} is added to all denominators.
\begin{lstlisting}[caption={Assembly of the demagnetization tensor $\boldsymbol{\tilde{N}}$. Only the six distinct components of the symmetric tensor are computed.},label=lst:demag_tensor]
def set_n_demag(c, permute, func):
  it = np.nditer(n_demag[:,:,:,c], flags=['multi_index'], op_flags=['writeonly'])
  while not it.finished:
    value = 0.0
    for i in np.rollaxis(np.indices((2,)*6), 0, 7).reshape(64, 6):
      idx=map(lambda k: (it.multi_index[k]+n[k]-1)%(2*n[k]-1)-n[k]+1,range(3))
      value+=(-1)**sum(i)*func(map(lambda j: (idx[j]+i[j]-i[j+3])*dx[j], permute))
    it[0]=-value/(4*pi*np.prod(dx))
    it.iternext()

n_demag = np.zeros([2*i-1 for i in n] + [6])
for i, t in enumerate(((f,0,1,2),(g,0,1,2),(g,0,2,1),
                       (f,1,2,0),(g,1,2,0),(f,2,0,1))):
  set_n_demag(i, t[1:], t[0])
\end{lstlisting}
The implementation of the tensor assembly is shown in listing~\ref{lst:demag_tensor}.
Basically the outer \texttt{while} loop iterates over the possible distances $\boldsymbol{i} - \boldsymbol{j}$ while the inner \texttt{for} loop corresponds to the sum in \eqref{eq:demag_n11} and \eqref{eq:demag_n12} respectively.
Within a regular grid of size $n_1 \times n_2 \times n_3$, possible index distances are $-n_x \leq i_x - j_x \leq n_x$.
Due to its symmetry only six components of the demagnetization tensor are computed.
This results in a total storage size of $2n_1-1 \times 2n_2-1 \times 2n_3-1 \times 6$ for the tensor.
\begin{figure}
  \centering
  \scalebox{0.8}{\includegraphics{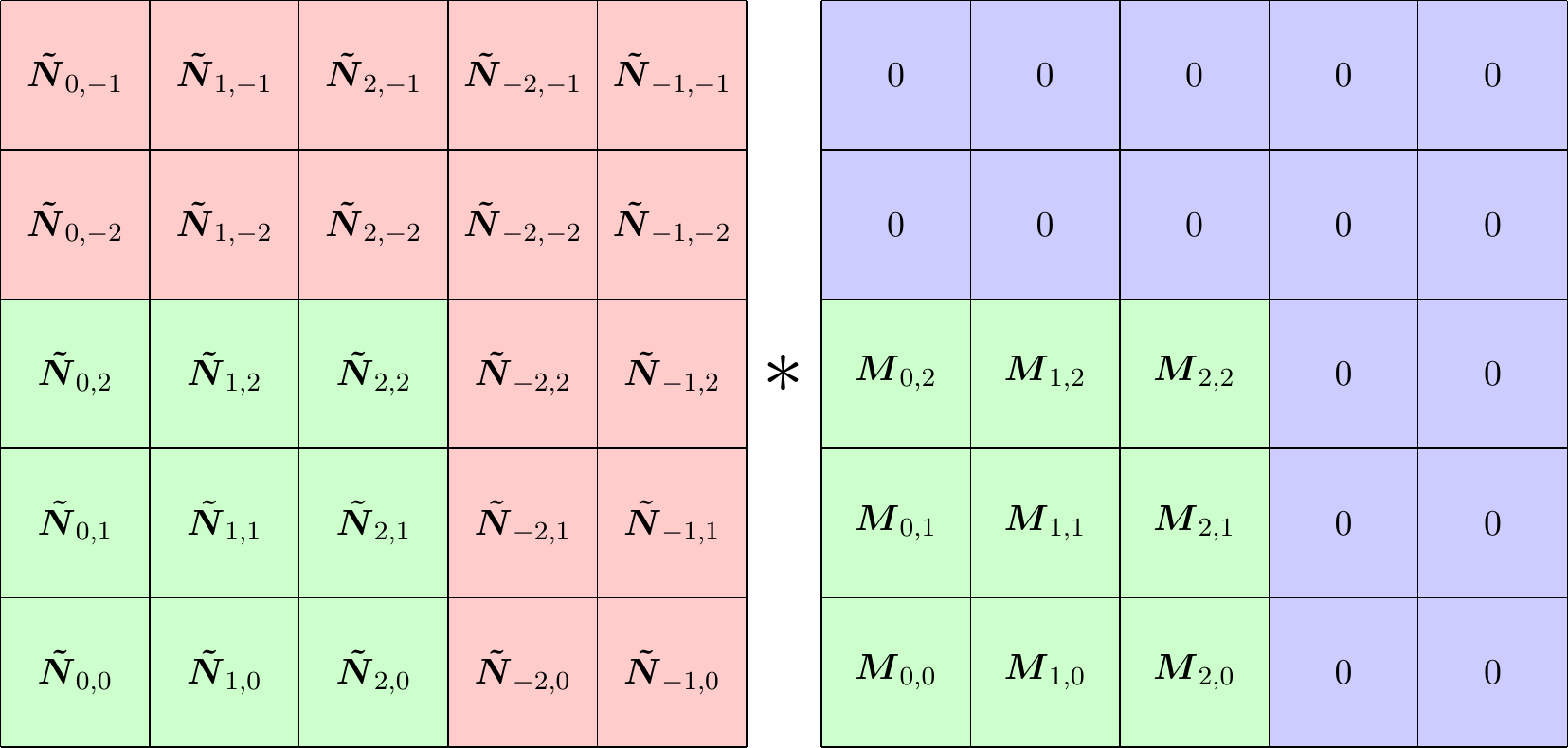}}
  \caption{
    Discrete convolution of the magnetization $\boldsymbol{m}$ with the demagnetization tensor $\boldsymbol{\tilde{N}}$.
    The numbering for the tensor entries start with zero and are wrapped around for negative distances.
    The magnetization is zero-padded to account for the cyclic nature of the fast convolution algorithm.
  }
  \label{fig:demag_convolution}
\end{figure}
Note that the spatial ordering of the tensor array is not done by increasing distance which would require the first element to hold the tensor for the largest negative distance.
Instead the numbering starts at zero distance and then cycles in a modulus fashion, see Fig.~\ref{fig:demag_convolution}.
This numbering is well suited for the application of the fast convolution as will be seen later.

The actual computation of the demagnetization field is performed in Fourier space where the discrete convolution in \eqref{eq:demag_convolution_discrete} reduces to a cell-wise multiplication
\begin{equation}
  \mathcal{F}(\boldsymbol{\tilde{N}} \ast \boldsymbol{m}) =
  \mathcal{F}(\boldsymbol{\tilde{N}}) \mathcal{F}(\boldsymbol{m}).
  \label{eq:convolution_theorem}
\end{equation}
This fast-convolution algorithm reduces the computational complexity of the demagnetization-field computation from $\mathcal{O}(N)$ for a naive implementaion of the convolution to $\mathcal{O}(N \log N)$ when using the fast Fourier transform algorithm.

Note that the fast convolution algorithm expects the convolution kernel to be of the same size as the function subject to the convolution in order to perform cell-wise multiplication in Fourier space.
The discrete magnetization however is of size $\prod_i n_i$ whereas the discrete tensor is of size $\prod_i 2n_i-1$.
Hence the discrete magnetization is expanded in order to match the size of the kernel, see Fig.~\ref{fig:demag_convolution}.
Note furthermore that the fast convolution applies the convolution kernel in a cyclic manner
\begin{equation}
  (f \ast g)_i = \sum_{j=0}^{n-1} = f_{(i-j+n)\%n} \cdot g_j
\end{equation}
, see \cite{press2007numerical}.
Thus the only reasonable choice for the expansion of the magnetization is the adding of zero entries, which is referred to as zero-padding.
This way, the computation of the field in a certain cell takes into account only contributions from the magnetization at physically possible cell distances.

\begin{lstlisting}[caption={Implementation of the fast convolution for the computation of the demagnetization field.},label=lst:demag_convolution]
m_pad = np.zeros([2*i-1 for i in n] + [3])
m_pad[:n[0],:n[1],:n[2],:] = m

f_n_demag = np.fft.fftn(n_demag, axes = filter(lambda i: n[i] > 1, range(3)))
f_m_pad   = np.fft.fftn(m_pad, axes = filter(lambda i: n[i] > 1, range(3)))

f_h_demag_pad = np.zeros(f_m_pad.shape, dtype=f_m_pad.dtype)
f_h_demag_pad[:,:,:,0] = (f_n_demag[:,:,:,(0, 1, 2)]*f_m_pad).sum(axis = 3)
f_h_demag_pad[:,:,:,1] = (f_n_demag[:,:,:,(1, 3, 4)]*f_m_pad).sum(axis = 3)
f_h_demag_pad[:,:,:,2] = (f_n_demag[:,:,:,(2, 4, 5)]*f_m_pad).sum(axis = 3)

h_demag = ms*np.fft.ifftn(f_h_demag_pad,
  axes = filter(lambda i: n[i] > 1, range(3)))[:n[0],:n[1],:n[2],:].real
\end{lstlisting}
The actual implementation of the fast convolution is shown in listing~\ref{lst:demag_convolution}.
The \texttt{fftn} and \texttt{ifftn} routines of NumPy perform an efficient three-dimensional Fourier transform.
The tensor-vector multiplication in Fourier space is implemented row-wise considering the symmetry of the tensor.

The presented code is a very basic implementation of a fast convolution algorithm.
Many improvements have been proposed in order to speed up this algorithm or to increase its accuracy.
A simple way to increase computation speed is the proper choice of the grid size.
Depending on the particular implementation, the fast Fourier transform performs better on certain list sizes.
In general most FFT implementations perform best on sizes that can be decomposed into small prime factors.
By additional zero padding the grid may be extended to a suitable size without changing the result of the computation.
A simple extension from $2n-1$ to $2n$ in every spatial dimension often leads to a noticable speedup.

Another performance gain may be achieved by ommitting Fourier transforms of zero entries.
These occur due to the zero-padding of the magnetization.
Likewise some inverse transforms may be omitted since only parts of the padded result are of physical meaning.
This procedure reduces the number of one-dimensional Fourier transforms to $7/12$, see \cite{kanai2010micromagnetic}.
By accounting for symmetries in the demagnetization tensor the algorithm can be further improved, see \cite{miltat2007numerical}.

Regarding accuracy the presented algorithm for the computation of the demagnetization tensor suffers from numerical problems especially for large distances and streched simulation cells.
These problems are usually overcome by switching between analytical formulas and numerical integration schemes depending on the cell distance, see \cite{donahueaccurate}.
\section{Exchange Field}\label{sec:exchange}
The exchange field models the quantum mechanical exchange interaction.
It is derived from a continuous formulation of the Heisenberg Hamiltonian and reads
\begin{equation}
  \boldsymbol{H}_\text{ex} = \frac{2 A}{\mu_0 M_\text{s}} \Delta \boldsymbol{m}.
  \label{eq:exchange_field}
\end{equation}
The second spatial derivative in the exchange field results in an additional boundary condition to the Landau-Lifshitz-Gilbert equation \eqref{eq:llg} in order to exhibit a unique solution.
It was shown in \cite{rado1959spin} that the Neumann condition
\begin{equation}
  \frac{\partial \boldsymbol{m}}{\partial \boldsymbol{n}} = 0
  \label{eq:neumann_boundary}
\end{equation}
has to hold if the boundary of the considered region is an outer boundary of a magnet.
The lowest order centered finite-difference approximation of the second derivative reads
\begin{equation}
  f''(x) \approx \frac{ f(x + \Delta x) - 2 f(x) + f(x - \Delta x)}{\Delta x^2}.
\end{equation}
With this, the three-dimensional Laplacian in \eqref{eq:exchange_field} can be approximated by
\begin{equation}
  \Delta \boldsymbol{m} (\boldsymbol{r}) \approx
  \sum_i \frac{
    \boldsymbol{m}(\boldsymbol{r} + \Delta r_i \boldsymbol{e}_i)
    - 2 \boldsymbol{m}(\boldsymbol{r})
    + \boldsymbol{m}(\boldsymbol{r} - \Delta r_i \boldsymbol{e}_i)
  }{\Delta r_i^2}
  \label{eq:exchange_fd}
\end{equation}
where $\Delta r_i$ is naturally chosen to match the cell size.
Not every simulation cell has neighboring cells in all six considered directions.
In case of a missing neighbor cell the magnetization of this cell is assumed to be the same as that of the center cell.
This procedure implicitly accounts for the Neumann boundary condition \eqref{eq:neumann_boundary}.
\begin{figure}
  \centering
  \scalebox{0.8}{\includegraphics{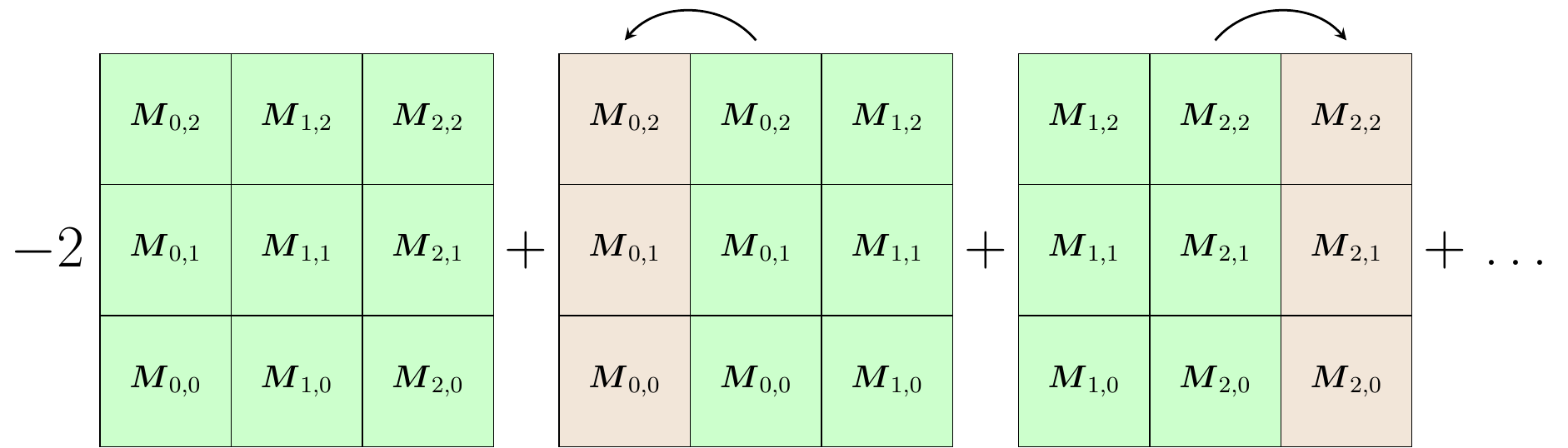}}
  \caption{
    Computation of the Laplacian with finite differences.
    The boundary values are duplicated in order to account for the zero Neumann boundary conditions.
  }
  \label{fig:exchange}
\end{figure}
\begin{lstlisting}[caption={Computation of the exchange field. The \texttt{repeat} method of NumPy is used to shift the magnetization while duplicating the boundary values.},label=lst:exchange]
h_ex = -2*m*sum([1/x**2 for x in dx])
for i in range(6):
  h_ex += np.repeat(m, 1 if n[i%3]==1 else \
                       [i/3*2]+[1]*(n[i%3]-2)+[2-i/3*2],axis=i%3)/dx[i%3]**2
h_ex *= 2*A/(mu0*ms)
\end{lstlisting}
Figure~\ref{fig:exchange} visualizes the implementation of \eqref{eq:exchange_fd}.
The finite differences for all cells are computed simultaneously by adding shifted versions of the arrays holding the magnetization values.
Listing~\ref{lst:exchange} shows the computation of the exchange field.

Possible improvements of this algorithm include the application of higher order finite-difference schemes.
However, the presented scheme turns out to deliver sufficient accurary for most problems and is implemented by many finite-difference codes.
\section{Landau-Lifshitz-Gilbert Equation}\label{sec:llg}
The numerical integration of the Landau-Lifshitz-Gilbert equation in the context of finite-difference micromagnetics is usually performed with explicit schemes.
Implicit schemes outperform explicit schemes in terms of stability which is especially useful in the case of stiff problems.
Due to the regular and well shaped grid, the stiffness of finite-difference problems is in general lower than that of finite-element methods on possibly ill shaped irregular meshes \cite{suess2002time}.
Also the stability of the implicit methods comes at the price of nonlinear problems due to the nonlinearity of the Landau-Lifshitz-Gilbert equation.

\begin{lstlisting}[caption={Projected explicit Euler scheme for the integration of the Landau-Lifshitz-Gilbert equation.},label=lst:llg]
dmdt = - gamma/(1+alpha**2) * np.cross(m, h) \
       - alpha*gamma/(1+alpha**2) * np.cross(m, np.cross(m, h))
m   += dt * dmdt
m   /= np.repeat(np.sqrt((m*m).sum(axis=3)), 3).reshape(m.shape)
\end{lstlisting}
The Landau-Lifshitz-Gilbert equation perserves the modulus of the magnetization at any point in space, i.e. $\partial |\boldsymbol{m}| / \partial t = 0$.
A simple integration scheme that accounts for this feature is a projected explicit Euler method
\begin{equation}
  \boldsymbol{m}(t + h) =
  \frac{\boldsymbol{m}(t) + h \partial_t \boldsymbol{m}(t)}
  {|\boldsymbol{m}(t) + h \partial_t \boldsymbol{m}(t)|}
\end{equation}
that is applied cellwise.
The implementation of this method is shown in listing~\ref{lst:llg}.

The performance of the time integration can be significantly improved by application of higher-order integration schemes.
A common choice in finite-difference micromagnetics is a Runge-Kutta scheme of 4th order with an adaptive step size based on a 5th order error estimation.

\section{Numerical Experiments}\label{sec:experiments}
\begin{figure}
  \centering
  \includegraphics{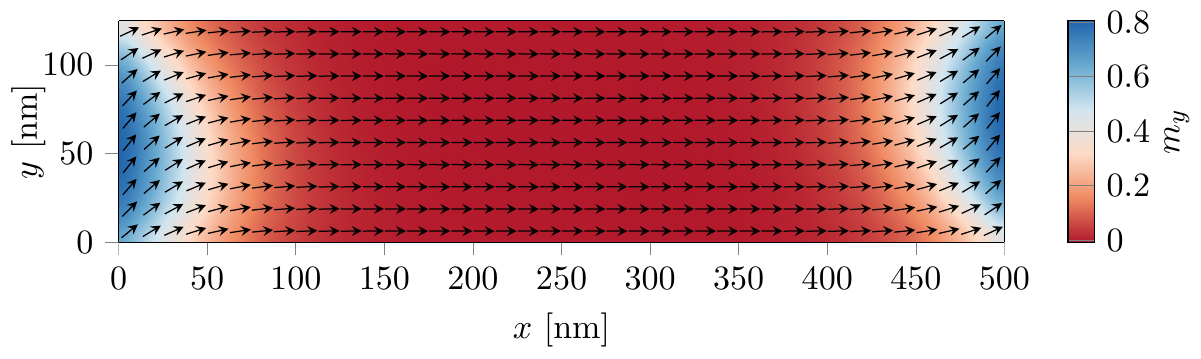}
  \caption{
    Magnetic s-state in a permalloy thin film as required as start configuration for the \textmu Mag standard problem \#4.
  }
  \label{fig:sstate}
\end{figure}
\begin{figure}
  \centering
  \includegraphics{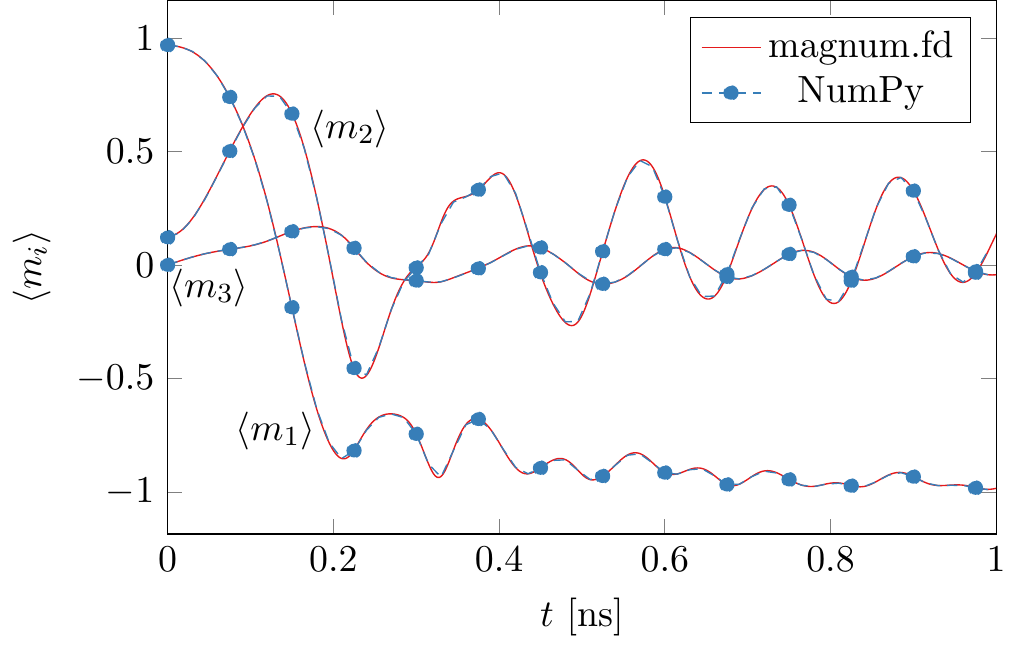}
  \caption{
    The time evolution of the averaged magnetization components $\langle m_i \rangle$ for the standard problem \#4.
  }
  \label{fig:sp4}
\end{figure}
In order to validate the implementation, the standard problem \#4 proposed by the \textmu Mag group is computed.
A magnetic cuboid with size $500 \times 125 \times 3\,\text{nm}$ and the material parameters of permalloy ($A = 1.3 \times 10^{-11} \,\text{J/m}$, $M_\text{s} = 8.0 \times 10^{5} \,\text{A/m}$, $\alpha = 0.02$) is considered.
The system is relaxed into a so-called s-state, see Fig.~\ref{fig:sstate}.
In a second step an external Zeeman field with a magnitude of $25\,\text{mT}$ is applied with an angle of $170^{\circ}$ to the sample.
The external field is directed opposite to the predominant magnetization direction and thus a switching process is initiated.
The simulation-cell size for this problem is chosen as $5 \times 5 \times 3\,\text{nm}$ and a time step of $h = 5\times10^{-15}\,\text{s}$ is used.
Figure~\ref{fig:sp4} shows the time evolution of the averaged magnetization components during switching.
The results are compared to a solution computed with the finite-difference code magnum.fd \cite{magnumfd} and show a good agreement.

\section{Conclusion}
We present a simple but complete micromagnetic finite-difference code written in less than 70 lines of Python.
While not competitive to mature micromagnetic simulation programs, the presented code provides all essential building blocks for complex micromagntic computations.
Due to the powerful and easy-to-use NumPy library it may serve as a prototyping environment for novel algorithms.
Furthermore the presented work is meant to give a brief introduction to finite-difference micromagnetics and provides references for detailed information on the topic.
A complete listing for the solution of the \textmu Mag standard problem \#4 is shown in the appendix and can be downloaded from \cite{70lines}.

\section*{Acknowledgements}
The financial support by
the Austrian Federal Ministry of Science, Research and Economy and the National Foundation for Research, Technology and Development
as well as
the FWF project SFB-ViCoM, F4112-N13
is gratefully acknowledged.

\begin{appendix}
\section{Code for Standard Problem \#4}
\begin{lstlisting}[]
import numpy as np
from math import asinh, atan, sqrt, pi

# setup mesh and material constants
n     = (100, 25, 1)
dx    = (5e-9, 5e-9, 3e-9)
mu0   = 4e-7 * pi
gamma = 2.211e5
ms    = 8e5
A     = 1.3e-11
alpha = 0.02

# a very small number
eps = 1e-18

# newell f
def f(p):
  x, y, z = abs(p[0]), abs(p[1]), abs(p[2])
  return + y/2.0*(z**2-x**2)*asinh(y/(sqrt(x**2+z**2)+eps)) \
         + z/2.0*(y**2-x**2)*asinh(z/(sqrt(x**2+y**2)+eps)) \
         - x*y*z*atan(y*z/(x * sqrt(x**2+y**2+z**2)+eps))   \
         + 1.0/6.0*(2*x**2-y**2-z**2)*sqrt(x**2+y**2+z**2)

# newell g
def g(p):
  x, y, z = p[0], p[1], abs(p[2])
  return + x*y*z*asinh(z/(sqrt(x**2+y**2)+eps))                 \
         + y/6.0*(3.0*z**2-y**2)*asinh(x/(sqrt(y**2+z**2)+eps)) \
         + x/6.0*(3.0*z**2-x**2)*asinh(y/(sqrt(x**2+z**2)+eps)) \
         - z**3/6.0  *atan(x*y/(z*sqrt(x**2+y**2+z**2)+eps))    \
         - z*y**2/2.0*atan(x*z/(y*sqrt(x**2+y**2+z**2)+eps))    \
         - z*x**2/2.0*atan(y*z/(x*sqrt(x**2+y**2+z**2)+eps))    \
         - x*y*sqrt(x**2+y**2+z**2)/3.0

# demag tensor setup
def set_n_demag(c, permute, func):
  it = np.nditer(n_demag[:,:,:,c], flags=['multi_index'], op_flags=['writeonly'])
  while not it.finished:
    value = 0.0
    for i in np.rollaxis(np.indices((2,)*6), 0, 7).reshape(64, 6):
      idx=map(lambda k: (it.multi_index[k]+n[k]-1)%(2*n[k]-1)-n[k]+1,range(3))
      value+=(-1)**sum(i)*func(map(lambda j: (idx[j]+i[j]-i[j+3])*dx[j],permute))
    it[0]=-value/(4*pi*np.prod(dx))
    it.iternext()

# compute effective field (demag + exchange)
def h_eff(m):
  # demag field
  m_pad[:n[0],:n[1],:n[2],:] = m
  f_m_pad = np.fft.fftn(m_pad, axes = filter(lambda i: n[i] > 1, range(3)))
  f_h_demag_pad = np.zeros(f_m_pad.shape, dtype=f_m_pad.dtype)
  f_h_demag_pad[:,:,:,0] = (f_n_demag[:,:,:,(0, 1, 2)]*f_m_pad).sum(axis = 3)
  f_h_demag_pad[:,:,:,1] = (f_n_demag[:,:,:,(1, 3, 4)]*f_m_pad).sum(axis = 3)
  f_h_demag_pad[:,:,:,2] = (f_n_demag[:,:,:,(2, 4, 5)]*f_m_pad).sum(axis = 3)

  h_demag = np.fft.ifftn(f_h_demag_pad,
    axes = filter(lambda i: n[i] > 1, range(3)))[:n[0],:n[1],:n[2],:].real

  # exchange field
  h_ex = - 2 * m * sum([1/x**2 for x in dx])
  for i in range(6):
    h_ex += np.repeat(m, 1 if n[i%3] == 1 else \
                         [i/3*2]+[1]*(n[i%3]-2)+[2-i/3*2], axis=i%3)/dx[i%3]**2

  return ms*h_demag + 2*A/(mu0*ms)*h_ex

# compute llg step with optional zeeman field
def llg(m, dt, h_zee = 0.0):
  h = h_eff(m) + h_zee
  dmdt = - gamma/(1+alpha**2) * np.cross(m, h) \
         - alpha*gamma/(1+alpha**2) * np.cross(m, np.cross(m, h))
  m += dt * dmdt
  return  m/np.repeat(np.sqrt((m*m).sum(axis=3)), 3).reshape(m.shape)

# setup demag tensor
n_demag = np.zeros([2*i-1 for i in n] + [6])
for i, t in enumerate(((f,0,1,2),(g,0,1,2),(g,0,2,1),
                       (f,1,2,0),(g,1,2,0),(f,2,0,1))):
  set_n_demag(i, t[1:], t[0])

m_pad     = np.zeros([2*i-1 for i in n] + [3])
f_n_demag = np.fft.fftn(n_demag, axes = filter(lambda i: n[i] > 1, range(3)))

# initialize magnetization that relaxes into s-state
m = np.zeros(n + (3,))
m[1:-1,:,:,0]   = 1.0
m[(-1,0),:,:,1] = 1.0

# relax
alpha = 0.50
for i in range(10000): llg(m, 5e-14)

# switch
alpha = 0.02
dt    = 5e-15
h_zee = np.tile([-24.6e-3/mu0, +4.3e-3/mu0, 0.0], np.prod(n)).reshape(m.shape)

with open('sp4.dat', 'w') as file:
  for i in range(int(1e-9/dt)):
    file.write("%f %f %f %f\n" % \
      ((i*1e9*dt,) + tuple(map(lambda i: np.mean(m[:,:,:,i]), range(3)))))
    llg(m, dt, h_zee)
\end{lstlisting}
\end{appendix}

\bibliographystyle{ieeetr}
\bibliography{refs}
\end{document}